\documentclass[12pt]{article}
\usepackage{setspace}
\usepackage{epsfig}
\usepackage{amssymb}
\usepackage{graphicx}
\def\be{\begin{equation}}
\def\ee{\end{equation}}
\def\ba{\begin{array}}
\def\ea{\end{array}}
\def\beqn{\begin{eqnarray}}
\def\eeqn{\end{eqnarray}}

\def\bt{\begin{tabular}}
\def\et{\end{tabular}}
\def\bc{\begin{center}}
\def\ec{\end{center}}
\begin{document}
\title{~~~Finding a unique texture for quark mass matrices}
\author{Samandeep Sharma, Priyanka Fakay, Gulsheen Ahuja$^*$, Manmohan
Gupta\\ {\it Department of Physics, Panjab University,
 Chandigarh, India.}\\
\\{\it $^*$gulsheen@pu.ac.in}}

\maketitle

\begin{abstract}
Within the Standard Model, starting with the most general mass
matrices we have used the facility of making weak basis
transformations and have imposed the condition of `naturalness' to
carry out their analysis within the texture zero approach.
Interestingly, our analysis reveals that a particular set of
texture 4 zero quark mass matrices can be considered to be a
unique viable option for the description of quark mixing data.
\end{abstract}

One of the key challenges in the present day high energy physics
is to understand vast spectrum of fermion masses and their
relationships with the corresponding mixing angles as well as mass
matrices. Despite impressive advances in the measurements of
fermion masses and mixing parameters, we are far from having a
compelling theory for flavor physics. Even for the case of quarks,
where precision measurements are available, the data is understood
in terms of phenomenological models having their roots in the
`bottom up' approach. In this context, exploring the possibility
of finding a minimal set of viable quark mass matrices can perhaps
be the first important step for solving the flavor riddle.

Without loss of generality, in the Standard Model (SM) and its
extensions where  right handed quarks are singlets, fermion mass
matrices can be considered to be hermitian using `Polar
Decomposition Theorem', for details we refer the reader to a
recent review by Gupta and Ahuja \cite{singreview}. The hermitian
matrices in the up and down sectors are characterized by 18 free
parameters which are still large in number as compared to the 10
observables i.e. 6 quark masses, 3 mixing angles and 1 CP
violating phase. In order that these matrices provide valuable
clues for developing theory of flavor physics, it is desirable
that following the bottom up approach the number of free
parameters of these are constrained by invoking certain broad and
general guidelines \cite{singreview,fxqrev}.

The bottom up approach of understanding fermion masses and mixings
has essentially evolved in three different directions. Firstly, on
the lines of Fritzsch ansatze \cite{frzans}, mass matrices are
formulated wherein certain elements of these are assumed to be
zero, usually referred to as texture zeros, and the compatibility
of the mixing matrix so obtained from these with the low energy
data ensures the viability of the formulated mass matrices. It has
been shown \cite{fxqrev,rrr} that both hermiticity and texture
zeros remain largely preserved while carrying out the
renormalization group evolution of the texture specific mass
matrices, therefore enabling one to formulate these at $M_Z$
scale. Despite showing considerable promise, in this approach the
possibility to arrive at a minimal set of viable quark mass
matrices emerges only by carrying out an exhaustive case by case
analysis of all possible texture zero mass matrices
\cite{singreview}.

Besides the above mentioned approach, within the framework of SM
and its extensions, one has the freedom to make unitary
transformations, referred to as `Weak Basis (WB) transformations',
which change the mass matrices without changing the quark mixing
matrix. Using WB transformations, several attempts
 \cite{fxwb}-\cite{costa} have been made wherein the above freedom
is exploited to introduce texture zeros in the quark mass
matrices. This results in somewhat reducing the number of free
parameters of general mass matrices, however, in the absence of
any constraints on the elements of the mass matrices, leads to a
large number of texture zero matrices which are able to explain
the quark mixing data \cite{costa}.

In yet another approach, advocated by Peccei and Wang  \cite{nmm},
the concept of `natural mass matrices' has been introduced to
formulate viable set of mass matrices at the Grand Unified
Theories (GUTs) as well as the $M_Z$ scale. In order to avoid fine
tuning, the elements of the mass matrices are constrained in order
to reproduce the hierarchical nature of the quark mixing angles.
This results in constraining the parameter space available to the
elements of the mass matrices, however without yielding a finite
set of viable mass matrices at the GUTs as well as the $M_Z$
scale.

A careful perusal of the above mentioned approaches suggests that
none of these lead to a finite set of viable texture specific mass
matrices, therefore in order to obtain the same perhaps one needs
to combine the three. The idea is to follow the texture zero
approach coupled with WB transformations to reduce the number of
free parameters of general hermitian mass matrices as well as to
impose the condition of `naturalness' for constraining the
parameter space available to the elements of these. The purpose of
the paper, therefore, is to start with the most general mass
matrices and consequently explore the possibility of obtaining a
finite set of viable texture specific mass matrices formulated by
using weak basis transformations as well as the constraints
imposed due to naturalness.

To begin with, we consider the following general hermitian mass
matrices
\be
 M_{q}=\left( \ba{ccc}
E_q & A _{q} & F _{q}     \\ A_{q}^{*} & D_{q} &  B_{U}     \\
 F _{q}^{*} & B_{q}^{*}  &  C_{q} \ea \right) ~~~~~~~~~~ (q=U,D),
 \label{genmm}
\ee
 which are related to the most general
mass matrices in the SM \cite{singreview}. As a next step, one can
introduce texture zeros in these matrices using the WB
transformations \cite{fxwb}, in particular, one can find a unitary
matrix $W$ transforming $M_U \rightarrow W^{\dagger} M_U W$ and
$M_D \rightarrow W^{\dagger} M_D W$, leading to
\be
 M_{U}=\left( \ba{ccc}
E_{U} & A _{U} & 0      \\ A_{U}^{*} & D_{U} &  B_{U}     \\
 0 &     B_{U}^{*}  &  C_{U} \ea \right), \qquad
M_{D}=\left( \ba{ccc} E_{D} & A _{D} & 0\\ A_{D}^{*} & D_{D} &
B_{D}     \\
 0 &     B_{D}^{*}  &  C_{D} \ea \right).
\label{t20}\ee  The above matrices, wherein $A_{q}= |A_{q}|e^{i
\alpha_{q}}$ and $B_{q}=|B_{q}|e^{i \beta_{q}}$ for $q=U, D$, can
be characterized as texture 2 zero quark mass matrices. It should
be noted that for $M_U$ and $M_D$ instead of zeros being in the
(1,3) and (3,1) positions, these could also be in either the (1,2)
and (2,1) or (2,3) and (3,2) position. These other structures are
related to the above mentioned matrices as we have the facility of
subjecting $M_U$ and $M_D$ to an another WB transformation which
can be the permutation matrix $P$. These different mass matrices,
however, yield the same mixing matrix, therefore while discussing
the results of the analysis, it is sufficient to discuss any one
of these matrices. Therefore, the matrices given in equation
(\ref{t20}) can now be considered as most general in the context
of SM. Further, in order to incorporate the condition of
`naturalness' on these mass matrices, the following condition is
imposed on the elements of the matrices \cite{unified}
\be
(1,i) < (2,j) \lesssim (3,3);~~~~~~~~~~~~~~~~~i=1,~2,~3, ~~j=2,~3.
\label{nathier} \ee After obtaining these matrices, as a next
step, their viability needs to be ensured by examining the
compatibility of the Cabibbo-Kobayashi-Maskawa (CKM) matrix
reproduced through these mass matrices with the recent quark
mixing data. This also enables one to check what constraints are
put on the elements of these mass matrices.

Before getting into the details of the analysis, we first present
some of the essentials pertaining to the construction of the CKM
matrix from these mass matrices. To facilitate diagonalization,
for $q=U,D$, the mass matrix $M_{q}$ may be expressed as $ M_q =
Q_q^{\dagger} M_q^r Q_q $ implying $ M^{r}_q = Q_q M_q
Q_q^{\dagger}$ where $ M^{r}_q$ is a symmetric matrix with real
eigenvalues and $ Q_q $ is the diagonal phase matrix, e.g., \be
M^{r}_{q} = \left(
 \ba {lll}
E_q &| A_{q}| & 0 \\ |A_{q}| & D_{q} &| B_{q}| \\ 0 & |B_{q}| &
C_{q}\ea \right), \qquad Q_{q}= \left( \ba  {ccc} e^{-i\alpha_{q}}
& 0 & 0 \\ 0 & 1 & 0 \\
            0 & 0 & e^{i\beta_{q}} \ea \right). \label{1fritzsch} \ee
The matrix $M_{q}^r$ can be diagonalized using the following
transformations
\be
M^{diag}_{q}= O^{T}_{q}M^{r}_{q}O_{q}= O^{T}_{q}Q_{q}
M_{q}Q^{\dagger}_q O_{q}= {\rm Diag}(m_{1}, -m_{2}, m_{3}), \ee
where the subscripts 1, 2 and 3 refer respectively to $u$, $c$,
$t$ for the up sector and $d$, $s$, $b$ for the down sector. The
exact diagonalizing transformation $O_q$ for the matrix $
M^{r}_{q}$ is given by
 \be O_q = \left(\ba{lll}
   {\sqrt \frac{(E_q+m_2)(m_3-E_q)(C_q-m_1)}{(C_q-E_q)(m_3-m_1)(m_2+m_1)}} &
    {\sqrt \frac{ (m_1-E_q)(m_3-E_q)(C_q+m_2)}{(C_q-E_q)(m_3+m_2)(m_2+m_1)}} &
  {\sqrt \frac{(m_1-E_q)(E_q+m_2)(m_3-C_q)}{(C_q-E_q)(m_3+m_2)(m_3-m_1)}}\\

  {\sqrt \frac{(C_q-m_1)(m_1-E_q)}{(m_3-m_1)(m_2+m_1)}} &
-{\sqrt \frac{(E_q+m_2)(C_q+m_2)}{(m_3+m_2)(m_2+m_1)}} &
 {\sqrt \frac{(m_3-E_q)(m_3-C_q) }{(m_3+m_2)(m_3-m_1)}} \\

  -{\sqrt \frac{(m_1-E_q)(m_3-C_q)(C_q+m_2)}{(C_q-E_q)(m_3-m_1)(m_2+m_1)}} &
 {\sqrt \frac{(E_q+m_2)(C_q-m_1)(m_3-C_q)}{(C_q-E_q)(m_3+m_2)(m_2+m_1)}} &
   {\sqrt \frac{(m_3-E_q)(C_q-m_1)(C_q+m_2)}{(C_q-E_q)
   (m_3+m_2)(m_3-m_1)}}  \ea \right). \label{oq} \ee
Further, these diagonalizing transformations  are related to the
mixing matrix as
\be
V_{\rm CKM} = O_{U}^{T} Q_{U} Q_{D}^{\dagger} O_{D}.
\label{ckmrel} \ee It should be noted that for the construction of
the CKM matrix, the elements $E_U$, $E_D$, $D_U$ and $D_D$ of the
mass matrices have been considered as free parameters.

The inputs used for the purpose of calculations, the quark masses
and the mass ratios at the $M_Z$ scale \cite{xing2012}, are
 \beqn m_u = 1.38^{+0.42}_{-0.41}\, {\rm MeV},~~~~~m_d =
2.82{\pm 0.48}\, {\rm MeV},~~~~ m_s=57^{+18}_{-12}\, {\rm
MeV},~~~\nonumber\\ m_c=0.638 ^{+0.043}_{-0.084}\, {\rm GeV},~~
m_b=2.86 ^{+0.16}_{-0.06}\, {\rm GeV},~~ m_t=172.1 {\pm 1.2} \,
{\rm GeV}, ~~~\\ \nonumber
 m_{u}/m_{d} = 0.553 {\pm 0.043} , m_{s}/m_{d} = 18.9 {\pm 0.8}
 .~~~~~~~~~~~~~~~~
\label{ratio}\eeqn The latest values \cite{pdg2012} of precisely
measured CKM parameters required for the construction of the CKM
matrix pertaining to three mixing angles and one CP violating
phase are \beqn |V_{us}| = 0.22534 \pm 0.00065,~~
|V_{ub}|=0.00351^{+0.00015}_{-0.00014}, ~~|V_{cb}|=
0.0412^{+0.0011}_{-0.0005}, \nonumber \eeqn \be {\rm Sin} 2\beta=
0.679 \pm 0.020.~~~~~~~~~~ \label{ckmvalues} \ee

Coming to the analysis of the mass matrices $M_U$ and $M_D$, the
parameters $\phi_1$ and $\phi_2$, related to the phases of the
mass matrices, $\phi_1$ = $\alpha_U- \alpha_D$ and $\phi_2=
\beta_U- \beta_D$, have been given full variation from 0 to
$2\pi$. Apart from $\phi_1$ and $\phi_2$, the free parameters
$E_U$, $E_D$, $D_U$ and $D_D$ have also been given wide variation
in conformity with the condition of naturalness as well as to
ensure that the elements of $O_U$ and $O_D$ should remain real. As
mentioned earlier, the viability of matrices $M_U$ and $M_D$ can
be checked by examining the compatibility of the CKM matrix so
reproduced, therefore, using the relation between mass matrices
and mixing matrix, given in equation (\ref{ckmrel}), the resultant
CKM matrix is obtained as follows
 \be
 {\rm V_ {CKM}}=\left( \ba{ccc}
0.9739-0.9745 & 0.2246-0.2259 & 0.00337-0.00365  \\ 0.2224-0.2259
& 0.9730-0.9990 &  0.0408-0.0422    \\ 0.0076-0.0101 &
0.0408-0.0422 &  0.9990-0.9999 \ea \right), \label{t20ckm}\ee
being fully compatible with the one given by PDG \cite{pdg2012}.
Also, the CP violating Jarlskog's rephasing invariant parameter
$J$ comes out to be $(2.494-3.365) \times 10^{-5}$ which again is
compatible with its latest experimental range, $(2.96
^{+0.20}_{-0.16})\times 10^{-5}$.

\begin{figure}[hbt]
\begin{minipage}{0.40\linewidth}   \centering
\includegraphics[width=2.0in,angle=-90]{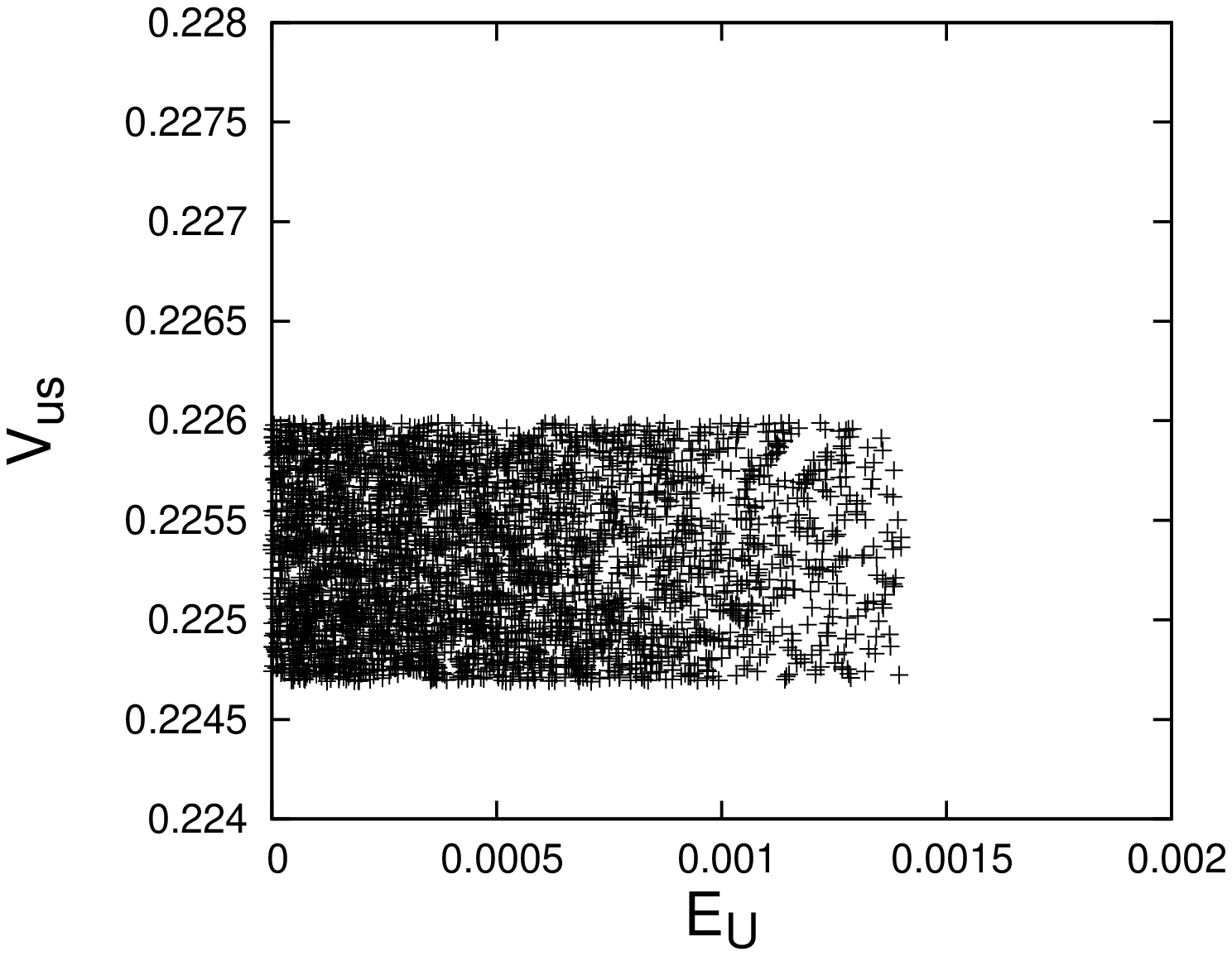}
\end{minipage} \hspace{0.5cm}
\begin{minipage} {0.40\linewidth} \centering
\includegraphics[width=2.0in,angle=-90]{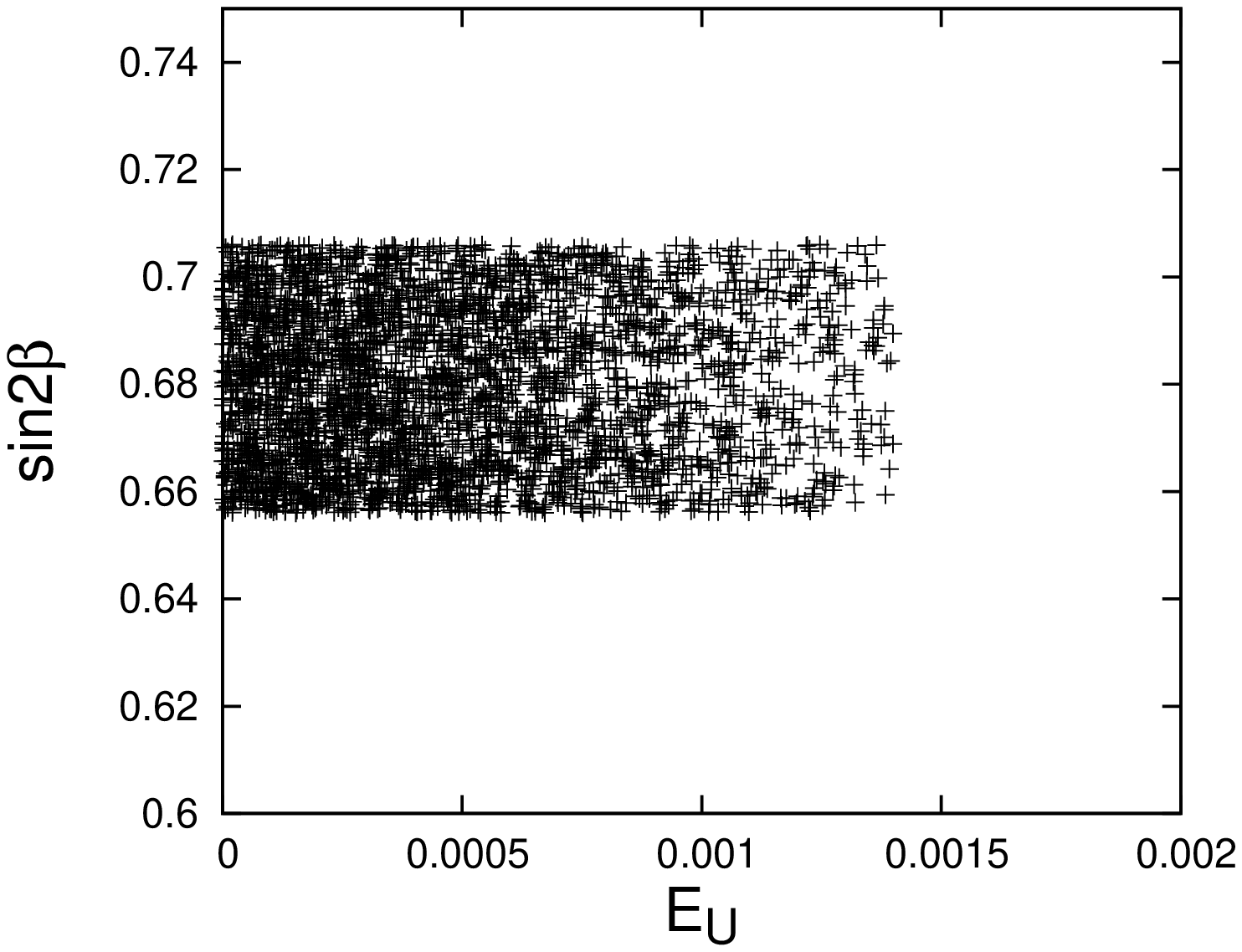}
 \end{minipage} \hspace{0.5cm}
\caption {Plots showing the dependence of $V_{us}$ and Sin2$\beta$
on the parameter $E_U$.} \label{fig1}
\end{figure}

The above mentioned compatibility leads to the viability of the
general mass matrices $M_U$ and $M_D$, however, since the number
of free parameters associated with these matrices is larger than
the number of observables, therefore, it becomes interesting to
examine whether any of their elements is redundant. To this end,
we present below the magnitudes of the elements of matrices $M_U$
and $M_D$ which reproduce the CKM matrix given in equation
(\ref{t20ckm})
\be
 M_{U}=\left( \ba{ccc}
0-0.00138 & 0.006-0.042 & 0     \\ 0.006-0.042 & 26.46-102.68 &
62.82-86.10\\ 0 & 62.82-86.10 & 68.78-145.00 \ea \right) {\rm
GeV}, \label{mu2z}\ee
 \be
M_{D}=\left( \ba{ccc} 0-0.00127 & 0.011-0.019 & 0     \\
0.011-0.019 & 0.36-1.66 & 1.03-1.44\\ 0 & 1.03-1.44 & 1.16-2.44
\ea \right) {\rm GeV}.\label{md2z} \ee A closer look at the above
matrices reveals that their (1,1) element is quite small in
comparison with the other non zero elements. This brings up the
issue whether the elements $E_U$ and $E_D$ of the matrices $M_U$
and $M_D$ respectively can be ignored all together without loss of
parameter space. To confirm this, one should examine the effect of
the variation of these parameters on CKM matrix elements. To this
end, in Figure (\ref{fig1}) we have plotted the variation of the
element $|V_{us}|$ and the CP asymmetry parameter Sin2$\beta$, two
of the best determined CKM parameters, with the mass matrix
element $E_U$. As is evident from these plots, not only for
reproducing the experimental ranges of $|V_{us}|$ and Sin2$\beta$,
mentioned in equation (\ref{ckmvalues}), the parameter $E_U$
assumes quite small values, $< 0.0014$ GeV, but also both
$|V_{us}|$ and Sin2$\beta$ seem independent of the range of $E_U$,
indicating the redundancy of element $E_U$. Similar conclusions
can be drawn from $E_U$ versus the other CKM matrix elements
plots. In the down sector, similar plots pertaining to (1,1)
element $E_D$ of the matrix $M_D$ reveal that again this parameter
is also quite small and essentially redundant.

These conclusions can be understood analytically also by examining
the exact transformation $O_{q}$, $q=U,D$, given in equation
(\ref{oq}). Interestingly, the parameters $E_U$ and $E_D$ being of
the order of the smallest mass eigenvalue $m_1$ can be ignored
when these appear with $m_2$, $m_3$ and $C_q$ which are much
larger than $m_1$. However, at other places $E_U$ and $E_D$ appear
with $m_1$ as $(m_1-E_q)$ which essentially implies re-scaling of
$m_1$.

Keeping in mind the above discussion, ignoring the elements $E_U$
and $E_D$ of the mass matrices, one gets $M_U$ and $M_D$ as
\be
 M_{U}=\left( \ba{ccc}
0 & A _{U} & 0      \\
A_{U}^{*} & D_{U} &  B_{U}     \\
 0 &     B_{U}^{*}  &  C_{U} \ea \right), \qquad
M_{D}=\left( \ba{ccc} 0 & A _{D} & 0\\
A_{D}^{*} & D_{D} & B_{D}     \\
 0 &     B_{D}^{*}  &  C_{D} \ea \right),
\label{flt40}\ee indicating a transition from texture 2 zero mass
matrices to texture 4 zero mass matrices. Carrying out a similar
analysis for these matrices, the corresponding CKM matrix comes
out to be
 \be
 {\rm V_ {CKM}}=\left( \ba{ccc}
0.9741-0.9744 & 0.2246-0.2259 & 0.00337-0.00365  \\ 0.2245-0.2258
& 0.9732-0.9736 &  0.0407-0.0422    \\ 0.0071-0.0100 &
0.0396-0.0417 &  0.9990-0.9992 \ea \right). \label{flckm}\ee This
matrix is not only in agreement with the latest quark mixing
matrix given by PDG \cite{pdg2012}, but is also fully compatible
with the CKM matrix given in equation (\ref{t20ckm}). Further, the
range of the CP violating Jarlskog's rephasing invariant parameter
$J$ comes out to be $(2.50-3.37) \times 10^{-5}$ which again is
compatible with its latest experimental range, justifying our
earlier conclusion that the elements $E_U$ and $E_D$ are
essentially redundant as far as reproducing the CKM parameters are
concerned.

It becomes interesting to examine how the parameter space of the
elements of the matrices $M_U$ and $M_D$ gets changed on going
from texture 2 to texture 4 zero. To this end, reconstructing
$M_U$ and $M_D$ we get
\be
 M_{U}=\left( \ba{ccc}
0 & 0.031-0.041 & 0     \\ 0.031-0.041 & 13.73-98.62 &
47.70-85.80\\ 0 & 47.70-85.80 &  72.84-157.73 \ea \right) {\rm
GeV},\label{mu4z} \ee
 \be
M_{D}=\left( \ba{ccc} 0 & 0.012-0.018  & 0     \\ 0.012-0.018 &
0.18-1.56 &  0.81-1.45     \\ 0 & 0.81-1.45  &  1.24-2.61 \ea
\right) {\rm GeV}. \label{md4z} \ee Interestingly, as expected,
the above matrices appear to be quite compatible with the earlier
mentioned matrices in equations (\ref{mu2z}) and (\ref{md2z}) and
the parameter space of the elements of the two also remains almost
the same, again confirming the redundancy of elements $E_U$ and
$E_D$.

\begin{table}
 \scalebox{0.6}{
\bt{|c|c|c|c|c|c|c|} \hline
 & a & b  & c & d & e & f  \\
 \hline &&&&&& \\
Category 1 & $\left ( \ba{ccc} {\bf 0} & A & {\bf 0} \\ A^{*}  & D
& B \\ {\bf 0} & B^{*}  & C \ea \right )$  &
 $\left ( \ba{ccc} {\bf 0} & {\bf 0}  & A \\ {\bf
0}  & C & B \\  A^{*} & B^{*}  & D \ea \right )$  &   $\left (
\ba{ccc} D & A &
  B\\ A^{*}  & {\bf 0}  & {\bf 0} \\  B^* & {\bf 0}
& C \ea \right )$ &  $\left ( \ba{ccc} C & B & {\bf 0}
 \\ B^{*} & D & A  \\ {\bf 0} & A^{*} & {\bf 0}
 \ea \right )$ &  $\left ( \ba{ccc} D & B & A
  \\ B^{*}  & C & {\bf 0} \\ A^{*} & {\bf 0}
& {\bf 0} \ea \right )$ & $\left ( \ba{ccc} C & {\bf 0} & B
 \\ {\bf 0}  & {\bf 0}  & A \\B^* & A^*  &
D \ea \right )$ \\ \hline &&&&&&\\

Category 2 & $\left ( \ba{ccc} D & A & {\bf 0} \\ A^{*}  & {\bf 0}
&  B \\ {\bf 0} & B^*  & C \ea \right )$     & $\left ( \ba{ccc} D
& {\bf 0} & A
 \\ {\bf 0} & C & B \\  A^* & B^*  &
{\bf 0} \ea \right )$  & $\left ( \ba{ccc} {\bf 0} & A &
 B \\ A^*  & D & {\bf 0}  \\  B & {\bf 0} &
C \ea \right )$ &  $\left ( \ba{ccc} C & B & {\bf 0} \\ B^*  &
{\bf 0} &  A \\ {\bf 0} & A^*  & D \ea \right )$   &  $\left (
\ba{ccc} C & {\bf 0} & B
 \\ {\bf 0} & D & A \\  B^* & A^*  &
{\bf 0} \ea \right )$
 & $\left ( \ba{ccc} {\bf 0} & B &
 A \\ B^*  & C & {\bf 0}  \\  A^*  & {\bf 0} &
D \ea \right )$ \\ \hline &&&&&&\\

Category 3  &$\left ( \ba{ccc} {\bf 0} & A &
 D\\ A^*  &{\bf 0}  & B \\ D^* & B^*  &
C \ea \right )$  &
  $\left ( \ba{ccc} {\bf 0} &
D & A
\\  D^*  & C  & B \\  A^* & B  &
 {\bf 0}\ea \right )$ & $\left ( \ba{ccc} {\bf 0} & A &
B
\\ A^* & {\bf 0} & D \\ B^*  & D^* &
C \ea \right )$ & $\left ( \ba{ccc} {\bf 0} & B &
 C\\ B^*  &{\bf 0}  & A \\ C^* & A^*  &
D \ea \right )$   & $\left ( \ba{ccc} {\bf 0} & C & B
\\  C^*  & D & A\\  B^* & A^*  &
 {\bf 0}\ea \right )$ & $\left ( \ba{ccc} {\bf 0} & B &
A
\\ B^* & {\bf 0} & C \\ A^*  & C^*
& D \ea \right )$    \\ \hline &&&&&&\\

Category 4 & $\left ( \ba{ccc} A & {\bf 0} & {\bf 0} \\ {\bf 0}  &
D & B\\ {\bf 0} & B^* & C \ea \right )$ & $\left (\ba{ccc}C & {\bf
0} & B
 \\ {\bf 0}  & A & {\bf 0} \\ B^*  & {\bf 0}  &
D \ea \right )$ & $\left ( \ba{ccc} C & B & {\bf 0} \\ B^*  & D &
{\bf 0} \\ {\bf 0} & {\bf 0}  & A \ea \right )$ &
    $\left ( \ba{ccc} A & {\bf 0} & {\bf 0} \\
{\bf 0}  & C & B\\ {\bf 0} & B^* & D \ea \right )$ & $\left
(\ba{ccc}D & {\bf 0} & B
 \\ {\bf 0}  & A & {\bf 0} \\ B^*  & {\bf 0}  &
C \ea \right )$  & $\left ( \ba{ccc} C & B & {\bf 0} \\ B^*  & D &
{\bf 0} \\ {\bf 0} & {\bf 0}  & A \ea \right )$\\
  \hline
\et \vspace{0.7 cm}} \caption{Table showing all possible texture 2
zero quark mass matrices, classified into four different
categories.} \label{3t4}
\end{table}

It may be noted that apart from the form of texture 4 zero mass
matrices considered above, there are several other possible
texture 4 zero structures \cite{singreview}. Based on whether the
matrices are related through permutations or not, all possible
texture 4 zero mass matrices can be classified as shown in Table
(\ref{3t4}). The matrices which are not related to each other
through permutations have been put into different categories. For
the matrices belonging to category 1, considering both $M_U$ and
$M_D$ as 1a type, corresponding to the ones mentioned in equation
(\ref{flt40}), we have already shown that these are viable and
explain the quark mixing data quite well. The other matrices of
this category, related through permutation matrix, also yield
similar results. For the matrices belonging to category 4, one
finds that interestingly these are not viable as in all these
matrices one of the generations gets decoupled from the other two.
Further, for categories 2 and 3, again a similar numerical
analysis reveals that the matrices of these classes are also not
viable as can be understood from the following CKM matrices
obtained for categories 2 and 3 respectively, e.g.,
\be
{\rm V_ {CKM}}=\left(\ba{ccc} 0.9740-0.9744 & 0.2247-0.2260 &
0.0024-0.0099\\ 0.2205-0.2256 & 0.9509-0.9727 & 0.0596-0.2172 \\
0.0140-0.0445 & 0.0584-0.2127 & 0.9905-1.0000\ea \right),
\label{4bckm} \ee
\be
{\rm V_ {CKM}}=\left(\ba{ccc} 0.9736-0.9744 &  0.2247-0.2260 &
0.0098-0.0331\\ 0.2226-0.2278 & 0.9549-0.9719 & 0.0659-0.1937\\
0.00007-0.0340 & 0.0694-0.1928 & 0.9810-0.9976\ea\right).
\label{5ackm} \ee A look at these matrices shows that there are
certain elements which do not agree with their corresponding
values given by PDG \cite{pdg2012}.

\begin{figure}[hbt]
\begin{minipage}{0.4\linewidth}   \centering
\includegraphics[width=1.8in,angle=-90]{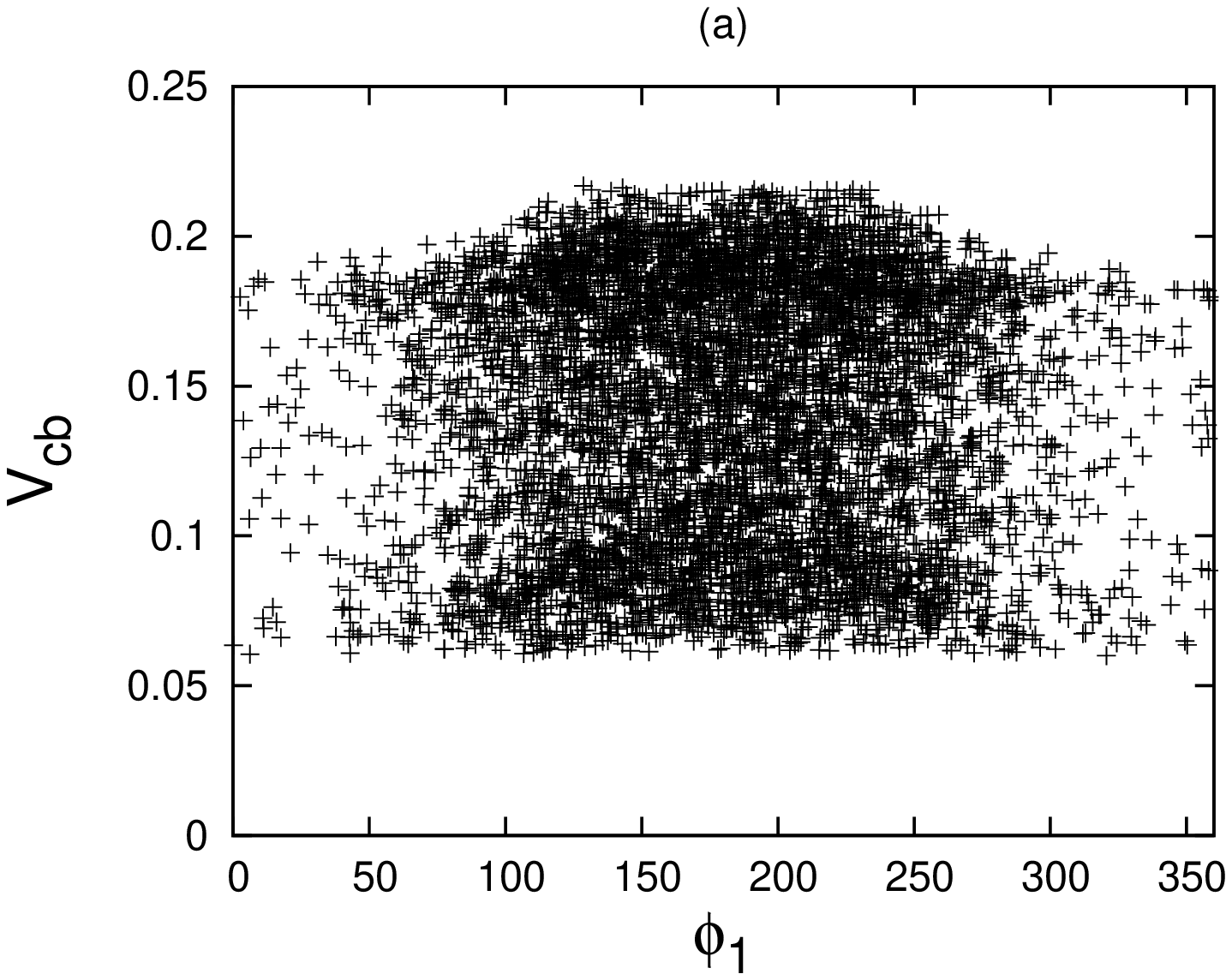}
\end{minipage} \hspace{0.5cm}
\begin{minipage} {0.4\linewidth} \centering
\includegraphics[width=1.8in,angle=-90]{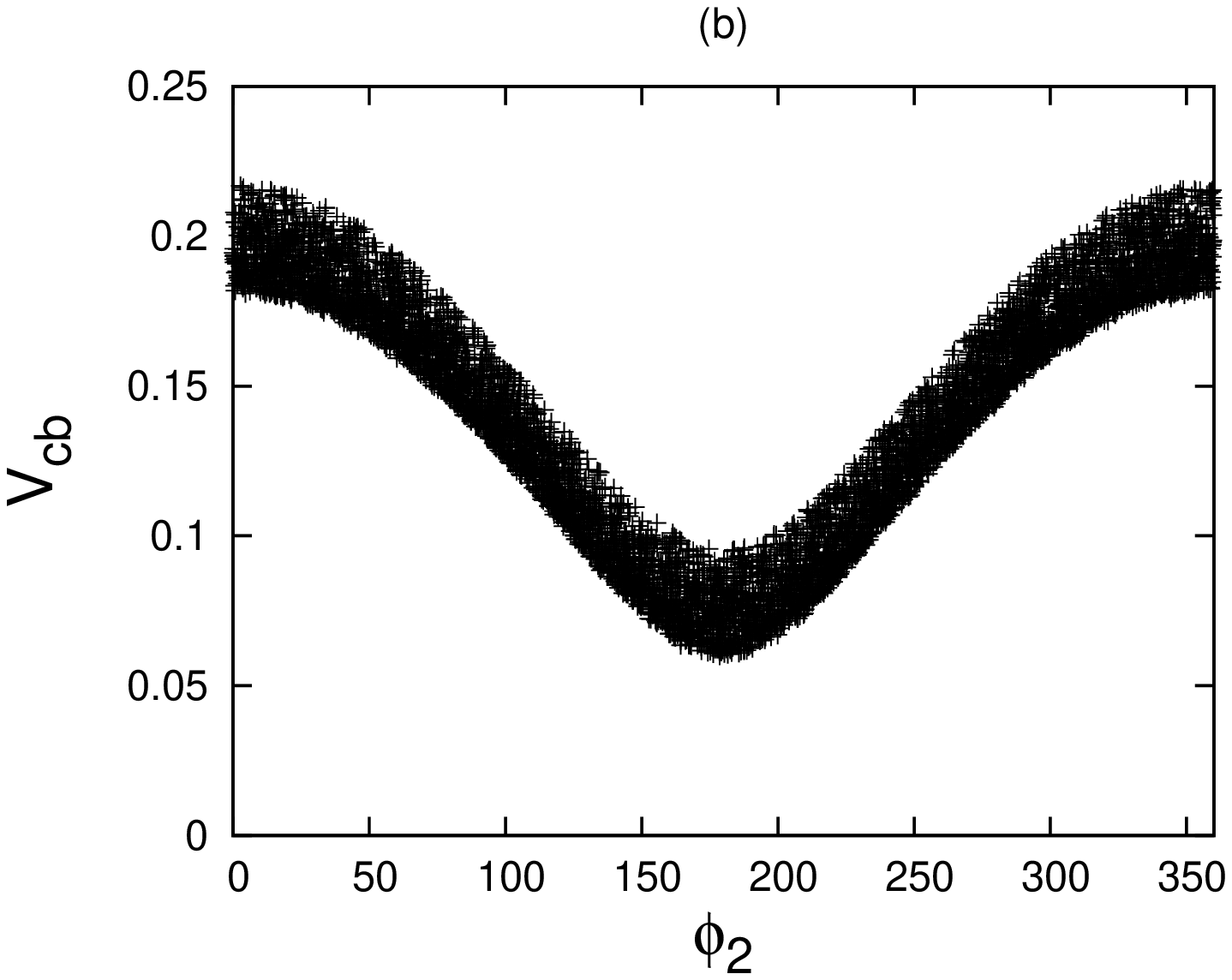}
 \end{minipage} \hspace{0.5cm}
 \\
 \\
 \\
\begin{minipage} {0.4\linewidth} \centering
\includegraphics[width=1.8in,angle=-90]{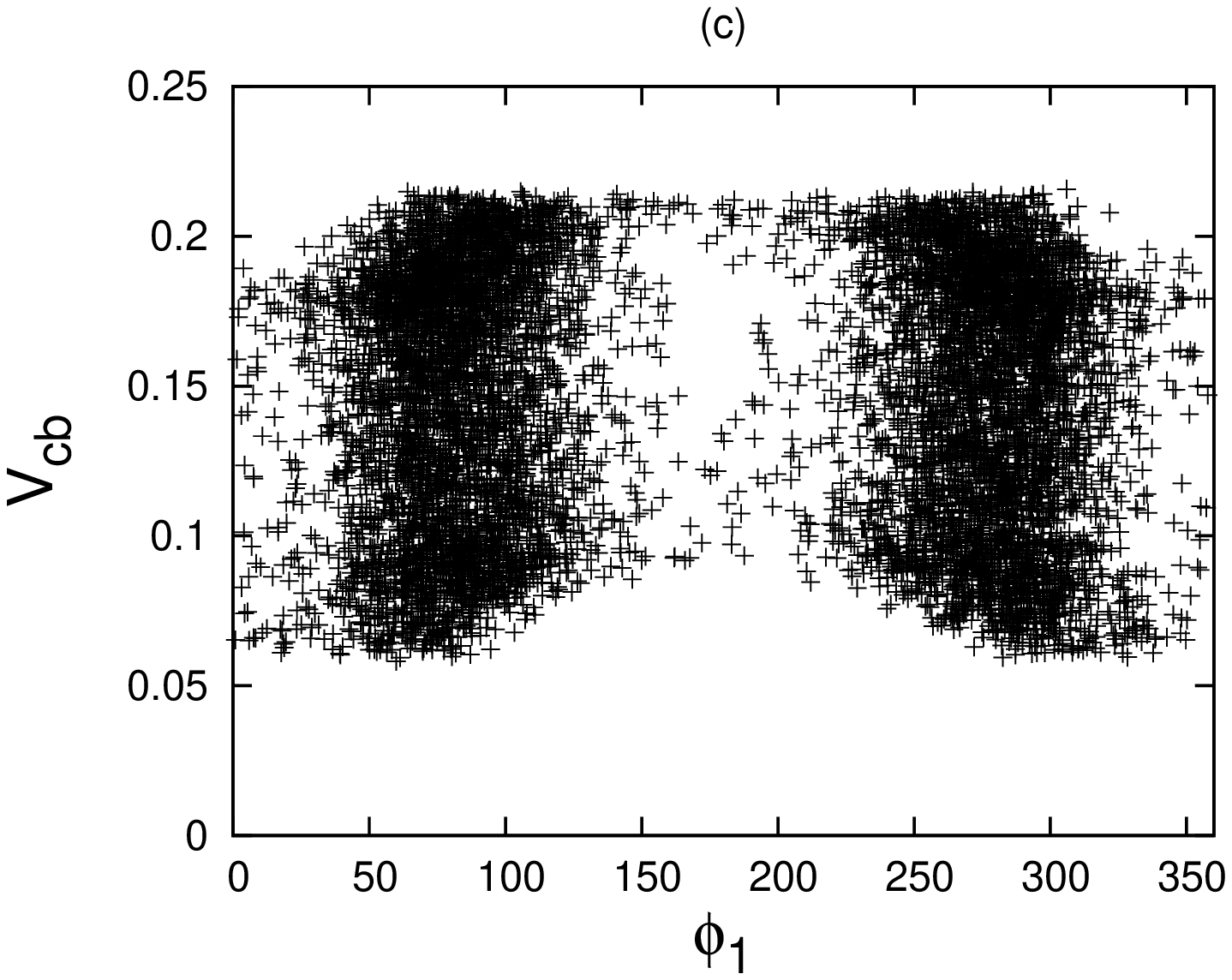}
 \end{minipage} \hspace{0.5cm}
 \begin{minipage} {0.4\linewidth} \centering
\includegraphics[width=1.8in,angle=-90]{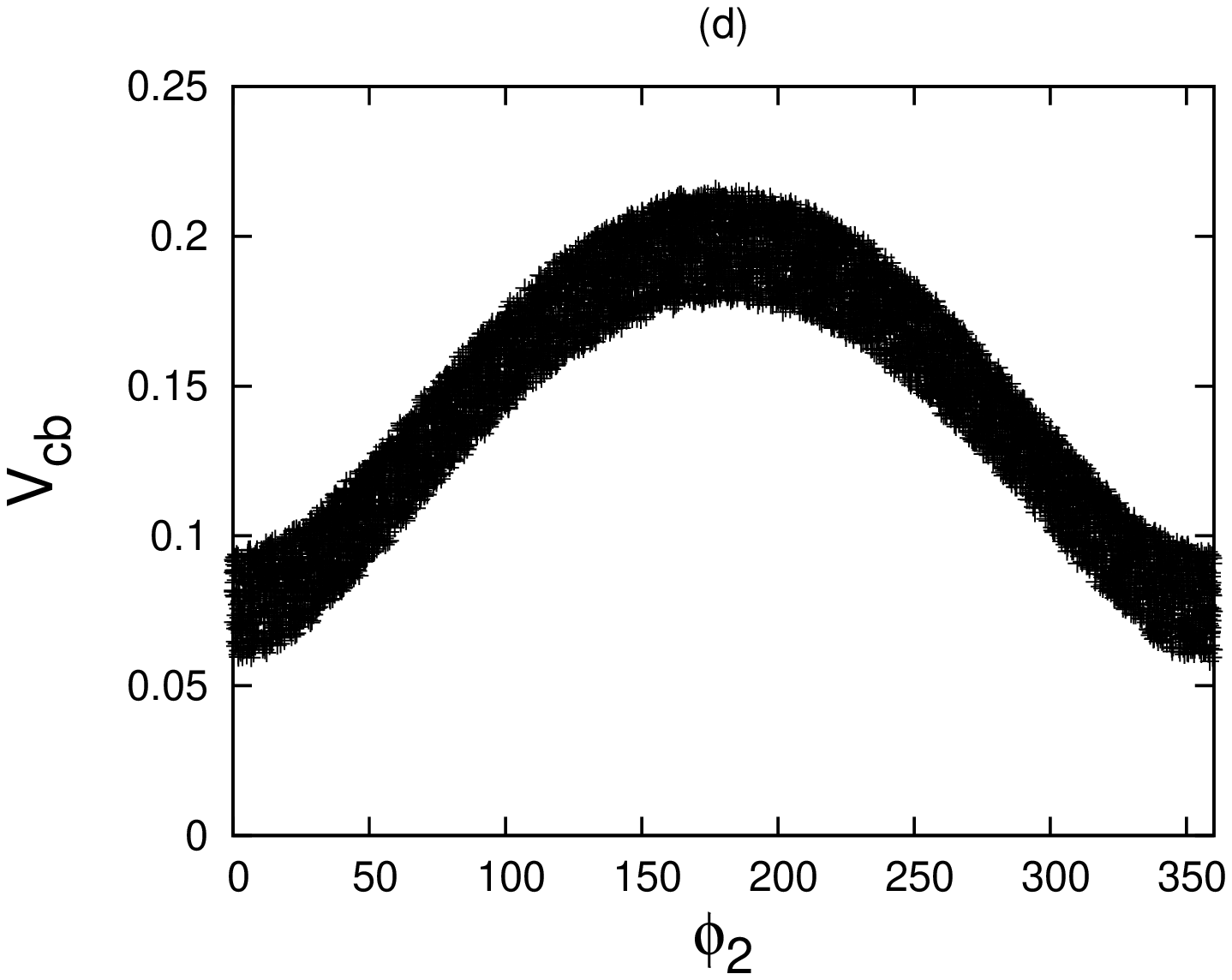}
 \end{minipage}
\caption {Plots showing the variation of the magnitude of $V_{cb}$
with phases $\phi_1$ and $\phi_2$ for quark mass matrices
belonging to categories 2 and 3 respectively.} \label{fig2}
\end{figure}

To check this rigorously, in Figure (\ref{fig2}) we have plotted
the precisely measured CKM matrix element $|V_{cb}|$ against the
phases $\phi_1$ and $\phi_2$ for the matrices belonging to
categories 2 and 3. While plotting these graphs, we have
constrained the value of element $V_{us}$ by its experimental
bounds given in equation (\ref{ckmvalues}), whereas full variation
has been given to the other parameters. From these graphs, one
finds that since the plotted values of element $|V_{cb}|$ have no
overlap with its experimental range, therefore, these matrices can
be considered to be non viable.

The above discussion clearly brings out that only the texture 4
zero quark mass matrices belonging to category 1 of the table are
found to be viable. Interestingly, the matrices given in equation
(\ref{flt40}) are quite similar to the original Fritzsch ansatze,
except for their (2,2) element being non zero for both $M_U$ and
$M_D$. In case one considers texture specific mass matrices with
zeros more than 4, we find that the present data rules out all
possible texture 5 and 6 zero quark mass matrices, confirming our
earlier conclusions in this regard \cite{singreview}.

To summarize, within the context of SM, starting with the most
general mass matrices, we have used the freedom of making WB
transformations to reduce these to texture 2 zero quark mass
matrices. Imposing the condition of `naturalness' within the
texture zero approach, one finds that certain elements of these
matrices can be considered as essentially redundant and therefore
reducing the matrices to texture 4 zero type. Numerical analysis
of all possible texture 4 zero mass matrices lead to a finite set
of these which can be considered as a unique viable option for the
description of quark mixing data . This texture structure for
quarks could be the first step towards unified textures for all
fermions.

\vskip 0.5cm {\bf Acknowledgements} \\ S.S. and G.A. respectively
thank UGC and DST, Government of India (Grant No:
SR/FTP/PS-017/2012) for financial support. S.S., P.F., G.A.
acknowledge the Chairperson, Department of Physics, P.U., for
providing facilities to work. M.G. would like to thank Walter
Grimus for useful discussions.


\begin{thebibliography}{99}
\bibitem{singreview} M. Gupta and G. Ahuja, Int. J. Mod. Phys. A {\bf 27},
1230033 (2012).

\bibitem{fxqrev} H. Fritzsch and Z. Z. Xing, Prog. Part. Nucl. Phys. {\bf 45}, 1 (2000).

\bibitem{frzans} H. Fritzsch, Phys. Lett. B {\bf 70}, 436
(1977); {\bf 73}, 317 (1978).

\bibitem{rrr} P. Ramond, R. G. Roberts and G. G. Ross, Nucl.
Phys. B {\bf 406}, 19 (1993).

\bibitem{fxwb} H. Fritzsch and Z. Z. Xing, Phys. Lett. B {\bf 413}, 396 (1997);
 Nucl. Phys. B {\bf 556}, 49 (1999).

\bibitem{brancowb}G. C. Branco, D. Emmanuel-Costa and R. Gonzalez Felipe, Phys. Lett. B {\bf 477}, 147
(2000); G.C. Branco, D. Emmanuel-Costa, R. Gonzalez Felipe and H.
Serodio, Phys. Lett. B {\bf 670}, 340 (2009).

\bibitem{costa} D. Emmanuel-Costa and C. Simoes, Phys. Rev. D {\bf 79}, 073006 (2009).

\bibitem{nmm} R. D. Peccei and K. Wang, Phys. Rev. D {\bf  53}, 2712 (1996).

\bibitem{unified}S. Sharma, P. Fakay, G. Ahuja and M. Gupta,
arXiv:1404.5726.

\bibitem{xing2012}Z.Z. Xing, H. Zhang and S. Zhou, Phys. Rev.
{\bf D  86}, 013013 (2012).

\bibitem{pdg2012}J. Beringer {\it et al.}, Particle Data Droup,
Phys. Rev. D {\bf 86}, 010001 (2012).

\end{thebibliography}
\end{document}